\documentclass[copyright]{eptcs}
\providecommand{\event}{DCFS 2010} % Name of the event you are submitting to
\usepackage{breakurl}             % Not needed if you use pdflatex only.

\title{Remembering Chandra Kintala}
\author{Martin Kappes
\institute{Faculty of Computer Science and Engineering\\
  University of Applied Sciences Frankfurt am Main, Germany}
\email{kappes@fb2.fh-frankfurt.de}
\and
Andreas Malcher
\institute{Institut f\"ur Informatik,
  Universit\"at Giessen\\
  Arndtstr.~2, 35392 Giessen, Germany}
\email{malcher@informatik.uni-giessen.de}
\and
Detlef Wotschke
\institute{Institut f\"ur Informatik,
  Goethe-Universit\"at Frankfurt\\
  60054 Frankfurt, Germany}
\institute{Institut f\"ur Informatik,
  Universit\"at Giessen\\
  Arndtstr.~2, 35392 Giessen, Germany}
\email{wotschke@em.uni-frankfurt.de}
}
\def\titlerunning{Remembering Chandra Kintala}
\def\authorrunning{M.~Kappes, A.~Malcher, D.~Wotschke}

\begin{document}%
\maketitle

\begin{abstract} With this contribution we would like to remember Chandra M. R. Kintala who
passed away in November 2009. We will give short overviews of his CV and his contributions 
to the field of theoretical and applied computer science and, given the opportunity, will 
attempt to present the current state of limited nondeterminism and limited resources for
machines. Finally, we will briefly touch on some research topics which hopefully will be
addressed in the not so distant future. 
\end{abstract}

\section{Introduction}\label{s:in}
On November 05, 2009, Chandra M. R. Kintala passed away at the age of 61, suddenly and unexpectedly. 

In this preliminary presentation and in a somewhat narrative, considerably more conceptual and less technical style, we would like to remember him as
a person, as a researcher and as a promoter and co-founder of descriptional complexity, the IFIP 
Working Group 1.2 and the workshops DCAGRS and DCFS. An extended, more comprehensive version in the
usual scientific style will follow.  

Section~\ref{s:cv} will briefly review Chandra's CV. In Section~\ref{s:tcs} we will focus on his direct scientific 
contributions to the area of limited resources and descriptional complexity, and in Section~\ref{s:spinoffs} we 
will continue with research areas and results which can be considered direct or indirect spin-offs
of his own research or which at least he promoted or which in their basic tenor resemble the flavor 
of his own approach to the area. Given this opportunity, we will attempt to present an overview of the current
state of the area of limited nondeterminism and limited resources for machines. In Section~\ref{s:practical}
we will give a brief survey of the many contributions 
Chandra made to the area of applied computer science and of some basic ideas for questions which in 
his mind theoreticians could formulate and investigate. Section~\ref{s:future} will purposely be of a somewhat 
speculative nature in that we will list some topics and general, but not specific and by no means
 well-defined, questions which in very recent discussions with him shortly before his death emanated 
as interesting and hopefully promising topics from the vantage point of descriptional complexity. 
We should conclude this introduction with the following disclaimer: Largely due to his continuous 
interaction with applied computer science, Chandra's view of limited resources and descriptional 
complexity was primarily influenced by a machine-based vantage point. For this reason and for the 
reason that we have to limit the scope of this contribution, we are taking a machine-based perspective 
and would like to apologize to everybody who otherwise should and would have been mentioned and cited.

\section{Chandra's CV}\label{s:cv}
Chandra Mohan Rao Kintala was born on July 22, 1948, in the city of Berhampur in the state of Orissa, 
a regional but sleepy hub on India's Eastern seaboard. He lost his father in his early teens, which 
made him face the reality of life quite early and which had a formative influence on him, both as a 
person and as a scientist. As a person in the sense that he always felt strong responsibility for 
everything he did and for the people who depended on him privately and professionally. As a scientist 
in the sense that, although he loved theoretical work, he never lost sight for the so-called more practical 
aspects and problems of computer science. He always saw theoretical and applied computer science as 
a unit, as two sides of the same coin. His ``credo'' can be summarized as follows: Theory should never 
lose sight of the practical questions and problems, and applied scientists should never stop explaining 
their problems to theoreticians and listen to what they have to say and offer. We will get back to this later. 

He received his Bachelor's Degree from Rourkela Engineering College and his Master's from IIT Kanpur. 
In 1973 he decided to continue his education in North America, first joined the PhD program at the University 
of Waterloo and then went with his advisor Patrick Fischer to the Pennsylvania State University in 1974 when 
Pat Fischer became Department Head there. In 1977 he received his PhD degree from Penn State with a thesis on limited nondeterminism of Turing Machines.

In 1977 he joined the University of Southern California as an Assistant Professor and worked with Seymour Ginsburg on 
Grammar Forms and with Ron Book, who was then at UC Santa Barbara, on questions in Formal Languages.

He married in 1978 and lived with his wife Bharati in Santa Monica, where their daughter Sreelata was born, until 1980. 
As much as Chandra enjoyed life there, professionally and privately, he missed a close interaction with the more practical
side of computer science. So when the opportunity arose to join Bell Labs in Murray Hill, he immediately saw this as
a chance and moved to New Jersey in 1980, where a few years later 
his son Kumar was born. 
Those years were, in his own words, among the happiest years of his life, as the research 
culture at Bell Labs provided exactly the close and stimulating interaction between theory and applications 
that he had always been looking for. And since he always liked teaching, he also taught courses at Steven's Institute 
of Technology in New Jersey, first as an Adjunct Professor and later as a Distinguished Industry Professor.

When Avaya Labs spun off from Bell Labs, he became Vice President of 
the Research Realization Center of Avaya Labs Research in Basking Ridge, New Jersey, where he continued the work he had begun at Bell Labs.

Although life, in his own words, was good to him, he faced the nagging question whether he should go back to his roots in 
India for at least a few years and how much he would still 
be able to blend in with the Indian culture and society. So, in Sept 2006, he went to Bangalore, India, to be the 
Director of Motorola Labs for two years and then joined Yahoo Labs for one year. There he was responsible for research in 
mobile communications, system sciences and for academic relations with top technical institutes in India. But 
during those three years he realized that he missed his two kids, who had stayed in the US, too much and that his real 
home had become New Jersey. So he moved back to the US and, in August 2009, joined the New Jersey Institute of Technology 
as Professor and Director of Software Engineering in its Computer Science Department.

In the late summer of 2009, only two months before his unexpected death, we had extensive talks with him. He was still 
full of ideas and had specific plans to investigate in a theoretical setting many of the practical computer science 
problems which he had run across during his time in industry. We will report on some of these ideas below in
Sections~\ref{s:practical} and~\ref{s:future}.

We would not do justice to the footprint which Chandra left on our community if we did not mention the instrumental role he played with 
others in founding the DCAGRS workshop,
which later turned into our current DCFS workshop. 
During his time at Bell Labs and later 
at Avaya Labs, Chandra became more and more 
involved with the problem of software reliability. 
And while the standard concept of software reliability 
often meant that software should run correctly and 
should only make  as few mistakes as possible, 
together with others he promoted the notion that 
correct and reliable software would also entail 
easily understandable and succinct descriptions 
of low descriptional complexity, where ``low'' 
would always have to be  defined within 
the respective applicable setting. So in 1998, together with others, 
he founded the workshop FDSR (Formal Descriptions and 
Software Reliability) as a satellite workshop to ISSRE 
(International Symposium on Software Reliability Engineering). 
As J{\"u}rgen Dassow pointed out at DCFS 2009 in Magdeburg, 
FDSR was essentially the beginning of our DCFS community and 
led to the first DCAGRS-DCFS workshop in Magdeburg in 
1999, as it was realized that descriptional complexity would 
be better off having its own focus, forum and workshop. 
Finally we should mention that he acted as Vice-Chair of our parent organization, the IFIP Working Group 1.2 for 
Descriptional Complexity, and took over as Chair when the position of Chair became vacant.

\section{Chandra's contributions to the area of theoretical computer science}\label{s:tcs}

Until the mid-seventies nondeterminism was, with a few exceptions, essentially viewed as a non-measur\-able or non-scalable resource. 
Either a machine was nondeterministic or it was not. Since early on Chandra saw theoretical computer science also through the prism 
of applied computer science, he realized that in practical models you often don't need the full or unlimited power of nondeterminism. 
A little amount of guessing is often all you need. Also, from a theoretical point, nondeterminism was at least for some automata classes 
such a powerful resource (some questions became computationally much more complex and in some instances even undecidable, when allowing 
full nondeterminism) that making nondeterminism a measurable and scalable resource should enable us to get a better handle on 
nondeterminism in general. So, in his PhD thesis and follow-up papers the concept of limited nondeterminism was introduced and investigated
for Turing Machine computations~\cite{FischerK78,KintalaF77,KintalaF80}. 

When Albert Meyer from MIT visited Penn State in 1975 and gave an inspiring talk on his and Michael Fischer's seminal paper on
the economy of description~\cite{MeyerF71}, it led to combining the concept of limited nondeterminism and limited resources with 
the concept of succinct descriptions, which was applied first to context-free languages in~\cite{Kintala78}. The concept of
limited nondeterminism was then applied to finite automata in~\cite{KintalaW80}, where it was shown that there is an infinite hierarchy of
finite amounts of nondeterminism generating an infinite hierarchy of trade-offs between NFAs and DFAs.

A particular way of limiting nondeterminism by weighing the various choices a machine can make and the resulting trade-offs to (unweighed) 
nondeterminism were investigated in the following papers on stochastic automata~\cite{KintalaPW93,KintalaW86}. 

Usually, when trying to show maximal trade-offs between
(finite state) machines of different degrees of 
nondeterminism, we construct 
specific witness languages which cause a particular trade-off,
and these languages change as the particular trade-offs change
that we 
want to achieve. In~\cite{GoldstineKW90} the concept of 
(nondeterminism)-spectra of regular languages was introduced 
in order to measure the \emph{inherent}
 nondeterminism of regular languages. 
The spectrum of a regular language is an infinite vector whose 
$i$th position lists the smallest number 
of states which any NFA with nondeterminism-degree $i$ needs to accept this language. As might be expected, 
there are regular languages which are very resilient
to finite nondeterminism so that any finite amount of nondeterminism does not suffice 
and where an infinite amount of nondeterminism is needed to yield a reduction in the number of states. And there are regular languages which are 
inherently very nondeterministic in the sense that every increase in the number of choices yields a reduction in the number of 
states until the smallest NFA has been obtained.

\section{Current state of the area of limited nondeterminism and limited resources for machines}\label{s:spinoffs}

Since its early days, the area of limited nondeterminism 
and limited resources for machines has come a long way, 
and although many of the papers published in this area since 
then have not been influenced by Chandra's work 
directly or indirectly, they resemble the flavor 
of his own approach to the area and are part of an area which, 
together with others, he started and promoted for many years. 
So we would like to take this opportunity and attempt to 
give an overview of the current state of this area. 
And again, as already done in the introduction, we would like to 
apologize for only looking at the machine-based perspective. 
Also, since even within this machine-based perspective 
alone the number of papers has become so enormous and since, due to space considerations, we have to limit the bibliography,
we would like to refer the reader to~\cite{Dassow09,GoldstineKKLMW02} for additional references. If, nevertheless,
we unintentionally might have overlooked a paper here or there 
which should have been included, please accept our apologies again. Moreover, since in this workshop contribution we
want to remember Chandra, the person as well as the theoretical \emph{and} applied 
scientist, this section will consist 
essentially only of a listing of the many and varied contributions to this area to date. In some cases,
strictly on a random basis and not emphasizing one area over another, we will include a few additional comments. 
A more detailed survey will have to be deferred to the extended version.

Limiting the full power of NFAs by considering 
subclasses of regular languages was studied for finite 
languages in~\cite{Mandl73,SalomaaY97}, for subregular languages in~\cite{BordihnHK09}, and for unary languages 
in~\cite{Chrobak86,Chrobak03,MereghettiP00}. The relationship 
between (limited) ambiguity and the size of NFAs was extensively 
treated in~\cite{GoldstineLW92,HromkovicS09,Leung98,Leung98-2,Leung05,Leung06,RavikumarI89,StearnsH85}. A somewhat 
different approach to limiting (or extending) nondeterminism by allowing multiple initial states (in DFAs and NFAs) was 
investigated in \cite{GillK74,HolzerSY01,Kappes99,VelosoG79}

Measuring nondeterminism based on communication complexity methods was dealt with in~\cite{HromkovicKKSS02}. 

The influence of nondeterminism and ambiguity on the size of B{\"u}chi automata was looked at in~\cite{Niessner04}. 

Turning to PDAs, non-recursive trade-offs were shown between unambiguous and deterministic PDAs in~\cite{Valiant76} and 
between ambiguous and unambiguous PDAs  in~\cite{SchmidtS77}. The concept of limited nondeterminism for PDAs was 
investigated in~\cite{GoldstineLW05,Herzog97,Herzog99,MrazPO07,SalomaaWY94,SalomaaY93,SalomaaY94,VermeirS81}, while 
limited ambiguity of PDAs or CFGs was studied in~\cite{Herzog97,Herzog99,Wich00,Wich01,Wich05}. 

Limiting the number of turns in successful computations of PDAs yields a hierarchy of non-recursive trade-offs with respect to the
number of turns~\cite{Malcher07}. These non-recursive trade-offs become ``only'' exponential and thus recursive, when bounded languages
are studied~\cite{MalcherP07}.

The limitations of nondeterminism mentioned so far concern the \emph{total amount} of nondeterminism applied, 
regardless of \emph{when} the nondeterminism occurs. When dealing with valuable and powerful resources it is often 
essential to use the resources not only economically but also in the
right situation and context or under appropriate conditions. This context-dependent use of nondeterminism was 
investigated for PDAs in~\cite{KutribM07a,KutribMW09,Masopust09},
whereas the context-dependent control of stack-turns in PDAs was studied in~\cite{KutribM07}. 

Limited resources for Turing Machines were investigated in~\cite{CaiC97,GoldsmithLM96,KleinK02,Kutrib02,Kutrib03}.

The effect of limiting resources on the computational capacity and the consequences for the descriptional complexity were also investigated
for the parallel model of cellular automata. For example, limited nondeterminism was considered in~\cite{BuchholzKK98,BuchholzKK02,BuchholzKK03}
and time hierarchies were refined in~\cite{BuchholzKK00,IwamotoHMI02}. Recently, limitations on the
number of cells~\cite{Malcher05,MalcherMP08} as well as on the amount of communication between the cells were studied in~\cite{KutribM09,KutribM09a}.

Finally, turning to parsers, in~\cite{GellerHSU77} an exponential trade-off was shown between arbitrary deterministic parsers, which are allowed to detect
an error arbitrarily late, and correct-prefix-parsers, which have to detect an error as early as possible. A ``limited approach'' to this trade-off 
was taken in~\cite{Fuessel92} by introducing and investigating various delay-spectra where the parsers become gradually smaller as the maximal delay 
in recognizing an error slowly increases. And finally looking at parsers for $\mathrm{LL}(k)$ and $\mathrm{LR}(k)$ languages, the effect on the 
parser-size by gradually increasing or decreasing the look-ahead was studied in~\cite{BertschN01,Blum01,LeungW00}.

\section{Chandra's contributions to the area of applied computer science}\label{s:practical}

Summarizing Chandra's almost 30 years of highly successful research and
management in industry is beyond the scope of this paper.
In his own words taken from Chandra's Website at
NJIT, ``Some of the noteworthy accomplishments (...) are

\begin{itemize}

\item a language and a software tool now called ``Backtalk'' with Dr.
David Belanger in 80s; it is still used in AT\&T for data analytics on
very large databases,

\item concepts and the components for Software-implemented Fault
Tolerance and Software Rejuvenation with Dr. Yennun Huang in early 90s;
 they are now widely used in industry and academia,

\item a Layer 7 network switch based on Linux for web content
distribution in his department in late 90s; it led to a new business
group in Lucent, and

\item an enterprise network monitoring system called Expertnet in early
2000s; it is still a key component in several products and services from
Avaya.''
\end{itemize}

Here, we will now focus on one select topic in which Chandra was
particularly active, namely Software Fault Tolerance. Please note that
for space reasons we decided to not include a detailed bibliography for
Chandra's applied work. Therefore we will also limit the references
for Software Fault Tolerance to some exemplary pointers only. 

Chandra was a
member of IFIP WG 10.4 on Dependable Computing and served as Program
Committee Member for Conferences such as ISSRE (International Symposium on
Software Reliability Engineering) and DSN (Dependable Systems and
Networks) for which he also was General Chair in 2006.  According to
Chandra's definition, Software Fault Tolerance is ``a set of software
components executing in the application layer of a computer system to
detect and recover from faults that are not handled in the underlying
hardware or operating system layers'' \cite{HuangK96}.

Software Fault Tolerance is an essential feature as more and more mission-critical 
tasks in enterprises are depending on the availability of services provided 
by IT systems. If a critical
application crashes or hangs, the consequences can be severe. Such a
software failure can have a number of reasons ranging from application
software faults to operating system software faults to faults in the
underlying hardware.

Many faults are transient, i.e., ``the failure may not recur if
the software is reexecuted on the same input'' \cite{HuangK96}.  For
example, consider a client/server application hanging because of
network-related issues or a failing application on a server due to an
unusual sequence of requests.

One of Chandra's most successful projects was SwiFT, Software
Implemented Fault Tolerance \cite{HuangK93}.
SwiFT used technologies such as
checkpointing and message logging \cite{WangHVCK95}. These terms denote
saving the state of a process onto a second machine or to storage at certain
intervals. In case of a fault, the process is stopped and recovered by
loading the last checkpointed state and replaying the logged messages
since then. The failure of a process is detected by a watchdog daemon
process which either periodically sends a signal to the application
process and waits for a response or receives heartbeat messages from
it. If the communication fails, the daemon waits for a certain period
and then retries. If the retry is also unsuccessful, it is assumed
that the application is hung and recovery is initiated.  Under
Chandra's supervision, such services were successfully implemented for
Unix and Windows NT \cite{LiangCHKLTW04} and are used in commercial products.

Furthermore, some components of SwiFT were extended to support
software rejuvenation, ``the concept of gracefully terminating an
application and immediately restarting it in a clean state''
\cite{HuangKKF95} in order to counter process aging due to, e.g.,
memory and/or file descriptor leaks. Software rejuvenation can be
beneficial in a variety of scenarios, see for instance
\cite{GargHKT96} or \cite{Kintala09}.

Clearly, the question whether a process is hung or not is undecidable
in the Turing sense. Therefore, more theoretically oriented approaches
and models for software reliability are based on other approaches such
as, e.g., statistical models like Markov Chains.
While Formal Languages and Automata Theory played
virtually no role in Software Fault Tolerance,
Chandra saw
 connections between these fields and hoped that, if the communities
worked closer together, Formal Languages and Automata could
provide valuable contributions to software reliability.

The question arises whether existing results on restarting
automata~\cite{JancarMPV95,Otto03,Otto06}
can be interpreted as a theoretical model for or answer
to the concept of rejuvenation or whether restarting automata can be
extended to model the concept of rejuvenation by, e.g., allowing the
automata, based on a certain strategy, to store states of the
computation which can be returned to when ``needed'' 
and from which the
 computation could resume.

Due to the complexity of large software systems and the cost involved
in building them, it is a common practice to use reusable commercial
off-the-shelf software components as building blocks.  In
\cite{KappesKK00}, Chandra and others presented a model for such systems
based on communicating finite state machines and showed that the
reliability of such a system cannot be precisely or approximately
calculated even if the reliability of the individual components is
known.  From a theoretical perspective (and with Rice's Theorem in
mind), the result might not seem surprising. However, Chandra considered
this paper as a first step to convince the software reliability
community of the usefulness and applicability of formal languages and
automata to problems in software reliability.

Chandra envisioned that studying the descriptional complexity of
new models tailored specifically at
software reliability could significantly advance our theoretical
knowledge in this area. His paper \cite{KappesK04} augmented his
efforts to convince the applied community of formal language models
with an approach to inspire the theoretical community to take a closer
look at software reliability in order to come up with new models of
interest to both the applied and the theoretical community. In his own
words,
``In order to start research relating descriptional complexity and
software reliability, first and foremost, models for software systems
need to be found and (possibly different) measures for their
``simplicity'' and their ``reliability'' have to be identified''
\cite{KappesK04}.

In this work, Chandra used finite automata as models for software
systems and studied trade-offs between reliability and conciseness of
such systems against an intended system behavior specified by a formal
language. ``A software system can be
considered a Turing Machine that processes a given input and produces
some output. Taking the abstraction one step further, a finite
automaton can be interpreted as strongly restricted version of a Turing
Machine. By restricting the output to ``yes'' or ``no'' only, the set
of all accepted inputs turns into a formal language and, along the
same lines, the expected behavior of the software system also becomes
a formal language. Much as the reliability of a software system needs
to be evaluated against its required functions, its specification, the
reliability of our model of the system, the DFA, needs to be evaluated
against a language specifying the intended system behavior''
\cite{KappesK04}.

Intuitively, the reliability of a finite automaton used as description
for the given specification is high if the difference between it and
the language accepted by the automaton is small. The presented results
included that the savings in the number of states between a fully
reliable and a less reliable representation cannot be bounded by any
function, even if the unreliable descriptions are required to exceed
any given fixed level of reliability. Moreover, \cite{KappesK04} shows
that, for a single regular language, there is a level of reliability
such that any description exceeding this level is at least as big as
the smallest DFA for the language. The quantitative measures in
\cite{KappesK04}  were later augmented by qualitative measures for the
reliability of Finite Automata in \cite{KappesN05}.

After returning to academia, Chandra had planned to resume his work to
bring experts from both communities together. One of the main
obstacles, in his view, was that problems of immense practical
interest often turn out to be intractable in theoretical models. While
a mathematical proof for, e.g., the undecidability of a problem in
a given model is a satisfactory and meaningful result for a
theoretician, a practitioner cannot simply put up with it. Even if the
problem is intractable, if it is relevant, it needs to be solved in one
way or another.

One of the ideas to cope with this situation is to start with simple yet
tractable models (such as Finite Automata) and then extend the models
towards more realistic scenarios while maintaining tractability.
Clearly, such a research project would require experts from the applied
as well as the theoretical community. With his
premature death, the task to pick up these
ideas is now left to others.

\section{Some possible future topics for descriptional complexity and limited resources}\label{s:future}

Since with this contribution we are remembering Chandra Kintala, we would like to list some ideas and ``visions''
for possible future topics in the area of descriptional complexity and limited resources. These ideas are the
result of discussions we had with him in the late Summer of 2009 when he had returned to New Jersey and, after
30 years in industry, had in his own words rediscovered his love and excitement for academia and theoretical
computer science in general and descriptional complexity in particular. The ideas we are going to list below
are of a ``visionary'' and speculative nature, are not (and cannot be yet) well-formulated or well-defined questions, 
are by no means fully thought through, might contain many pitfalls or completely intractable hurdles, and they might 
actually be controversial in the sense whether one should move in this direction at all. In other words, these ideas 
ought to be taken as suggestions for topics and not as well-formulated research questions. And, as mentioned above, 
all this can only be very sketchy here and by no means complete.

As intriguing, fascinating and often difficult (to prove) non-recursive trade-offs are, from the perspective of an
applied computer scientist they are not very useful, except for the insight that one might step onto dangerous territory.
For, as much as one is interested in maximal compressions, these compressions should still contain enough information to
algorithmically restore descriptions to their original state. In other words, our non-recursive hierarchies, 
e.g., when increasing the degree of nondeterminism or ambiguity or the number of turns in PDAs, are still too coarse. 
This might be caused by the fact that we are considering PDAs where the stack heights, or even more so, the stack 
contents of the various computations for an input are not ``structurally'' related, e.g., not related by a recursive
function. Let us bear in mind that for accepting the invalid computations of a Turing Machine we use nondeterministic
PDAs where the different computations on a given input show ``opposite'' stack behavior. What effect on the trade-offs
in descriptional complexity would a restriction have that requires the stack-contents or stack heights to be
structurally or functionally related by some recursive function? Would we then get infinite recursive hierarchies of
 trade-offs which are based on the recursive functions or structural relations used? Would such restrictions be useful from an applied perspective? Probably more than non-recursive trade-offs? 

When we prove (infinite hierarchies of) non-recursive trade-offs, we usually use witness languages which account for a 
specific jump from one level to the next within those hierarchies. Again, as enlightening as this might be from a theoretical 
vantage point, it is not exactly what applied computer scientists are confronted with. They are normally faced with a
specific language which to elegantly describe they have several tools (resources) of varying degrees at their disposal.
So questions arise like: How much of a certain resource can or should one apply in order to get the description down to a
reasonable size, if that is at all possible? Does the spectrum of the language  (cf.~\cite{GoldstineKW90}) generated by ever increasing amounts
of a certain resource show a ``steady decline'' in size or does it remain constant up to a point from which on it collapses ``dramatically''? 
These ideas lead us to suggest to investigate, e.g., for context-free languages, ambiguity-spectra, nondeterminism-spectra, 
turn-spectra, state-spectra and other spectra of limited resources. 

In his practical work, Chandra often had to work with attribute grammars, e.g.~\cite{Kintala83}, which were introduced by
Knuth~\cite{Knuth68} in order to obtain elegant context-free descriptions, e.g., of binary numbers. These grammars work with 
inherited and synthesized attributes, and often small or elegant descriptions can be obtained by carefully balancing the 
interaction of these different kinds of attributes. What happens to the size of the grammar when we gradually shift away 
from the optimal balance by limiting the resource of inherited attributes in favor of synthesized attributes or vice versa? A first step in this direction was undertaken in~\cite{Sunckel98}. 

The above list of possible topics is only a sample list of topics which lie at the juncture of applied computer science and 
descriptional complexity and reflect Chandra's view of our field. More on all this in an extended version.

\bibliographystyle{eptcs} % or whatever you prefer

\begin{thebibliography}{10}

\bibitem{BertschN01}
Eberhard Bertsch and Mark-Jan Nederhof.
\newblock Size/lookahead tradeoff for {L}{L}$(k)$-grammars.
\newblock {\em Inform. Process. Lett.}, 80(3):125--129, 2001.

\bibitem{Blum01}
Norbert Blum.
\newblock On parsing {L}{L}-languages.
\newblock {\em Theoret. Comput. Sci.}, 267(1--2):49--59, 2001.

\bibitem{BordihnHK09}
Henning Bordihn, Markus Holzer, and Martin Kutrib.
\newblock Determination of finite automata accepting subregular languages.
\newblock {\em Theor. Comput. Sci.}, 410(35):3209--3222, 2009.

\bibitem{BuchholzKK98}
Thomas Buchholz, Andreas Klein, and Martin Kutrib.
\newblock One guess one-way cellular arrays.
\newblock In {\em Mathematical Foundations of Computer Science ({MFCS} 2000)},
  volume 1450 of {\em Lecture Notes in Comput. Sci.}, pages 807--815.
  Springer-Verlag, 1998.

\bibitem{BuchholzKK00}
Thomas Buchholz, Andreas Klein, and Martin Kutrib.
\newblock Iterative arrays with small time bounds.
\newblock In {\em Mathematical Foundations of Computer Science ({MFCS} 2000)},
  volume 1893 of {\em Lecture Notes in Comput. Sci.}, pages 243--252.
  Springer-Verlag, 2000.

\bibitem{BuchholzKK02}
Thomas Buchholz, Andreas Klein, and Martin Kutrib.
\newblock On interacting automata with limited nondeterminism.
\newblock {\em Fund. Inform.}, 52(1--3):15--38, 2002.

\bibitem{BuchholzKK03}
Thomas Buchholz, Andreas Klein, and Martin Kutrib.
\newblock Iterative arrays with limited nondeterministic communication cell.
\newblock In {\em Words, Languages and Combinatorics III}, pages 73--87. World
  Scientific Publishing, 2003.

\bibitem{CaiC97}
Liming Cai and Jianer Chen.
\newblock On the amount of nondeterminism and the power of verifying.
\newblock {\em SIAM J. Comput.}, 26(3):733--750, 1997.

\bibitem{Chrobak86}
Marek Chrobak.
\newblock Finite automata and unary languages.
\newblock {\em Theoret. Comput. Sci.}, 47(2):149--158, 1986.

\bibitem{Chrobak03}
Marek Chrobak.
\newblock Errata to: `{F}inite automata and unary languages'.
\newblock {\em Theoret. Comput. Sci.}, 302(1--3):497--498, 2003.

\bibitem{Dassow09}
J{\"u}rgen Dassow.
\newblock Ten years {DCFS}.
\newblock Presentation at DCFS 2009, Magdeburg, 2009. Electronic resource {\tt
  http://theo.cs.uni-magdeburg.de/dcfs2009/text/TenYearsDCFS.pdf}.

\bibitem{FischerK78}
Patrick~C. Fischer and Chandra M.~R. Kintala.
\newblock Real-time computations with restricted nondeterminism.
\newblock {\em Math. Systems Theory}, 12(3):219--231, 1978/79.

\bibitem{Fuessel92}
Hans-Martin F{\"u}ssel.
\newblock Komplexit{\"a}t einer fr{\"u}hen {F}ehlererkennung beim {P}arsen von
  deterministisch kontextfreien {S}prachen.
\newblock Master's thesis, J.W. Goethe-Universit{\"a}t Frankfurt, 1992.

\bibitem{GargHKT96}
Sachin Garg, Yennun Huang, Chandra M.~R. Kintala, and Kishor~S. Trivedi.
\newblock Minimizing completion time of a program by checkpointing and
  rejuvenation.
\newblock In {\em ACM SIGMETRICS Conference on Measurement and Modeling of
  Computer Systems}, pages 252--261, 1996.

\bibitem{GellerHSU77}
Matthew~M. Geller, Harry~B. Hunt, III, Thomas~G. Szymanski, and Jeffrey~D.
  Ullman.
\newblock Economy of description of parsers, {D}{P}{D}{A}'s, and {P}{D}{A}'s.
\newblock {\em Theoret. Comput. Sci.}, 4(2):143--153, 1977.

\bibitem{GillK74}
Arthur Gill and Lawrence~T. Kou.
\newblock Multiple-entry finite automata.
\newblock {\em J. Comput. System Sci.}, 9:1--19, 1974.

\bibitem{GoldsmithLM96}
Judy Goldsmith, Matthew~A. Levy, and Martin Mundhenk.
\newblock Limited nondeterminism.
\newblock {\em SIGACT News}, 27(2):20--29, 1996.

\bibitem{GoldstineKKLMW02}
Jonathan Goldstine, Martin Kappes, Chandra M.~R. Kintala, Hing Leung, Andreas
  Malcher, and Detlef Wotschke.
\newblock Descriptional complexity of machines with limited resources.
\newblock {\em J.UCS}, 8(2):193--234, 2002.

\bibitem{GoldstineKW90}
Jonathan Goldstine, Chandra M.~R. Kintala, and Detlef Wotschke.
\newblock On measuring nondeterminism in regular languages.
\newblock {\em Inform. and Comput.}, 86(2):179--194, 1990.

\bibitem{GoldstineLW92}
Jonathan Goldstine, Hing Leung, and Detlef Wotschke.
\newblock On the relation between ambiguity and nondeterminism in finite
  automata.
\newblock {\em Inform. and Comput.}, 100(2):261--270, 1992.

\bibitem{GoldstineLW05}
Jonathan Goldstine, Hing Leung, and Detlef Wotschke.
\newblock Measuring nondeterminism in pushdown automata.
\newblock {\em J. Comput. System Sci.}, 71(4):440--466, 2005.

\bibitem{Herzog97}
Christian Herzog.
\newblock Pushdown automata with bounded nondeterminism and bounded ambiguity.
\newblock {\em Theoret. Comput. Sci.}, 181(1):141--157, 1997.

\bibitem{Herzog99}
Christian Herzog.
\newblock {\em Die Rolle des Nichtdeterminismus in kontextfreien Sprachen}.
\newblock PhD thesis, Johann Wolfgang Goethe-Universit{\"a}t, Frankfurt, 1999.

\bibitem{HolzerSY01}
Markus Holzer, Kai Salomaa, and Sheng Yu.
\newblock On the state complexity of $k$-entry deterministic finite automata.
\newblock {\em J. Autom. Lang. Comb.}, 6(4):453--466, 2001.

\bibitem{HromkovicKKSS02}
Juraj Hromkovi{\v{c}}, Juhani Karhum{\"a}ki, Hartmut Klauck, Georg Schnitger,
  and Sebastian Seibert.
\newblock Communication complexity method for measuring nondeterminism in
  finite automata.
\newblock {\em Inform. and Comput.}, 172(2):202--217, 2002.

\bibitem{HromkovicS09}
Juraj Hromkovi{\v{c}} and Georg Schnitger.
\newblock Ambiguity and communication.
\newblock In {\em Symposium on Theoretical Aspects of Computer Science ({STACS}
  2009)}, volume~3 of {\em LIPIcs}, pages 553--564. Schloss Dagstuhl -
  Leibniz-Zentrum f{\"u}r Informatik, Germany, 2009.

\bibitem{HuangK93}
Yennun Huang and Chandra M.~R. Kintala.
\newblock Software implemented fault tolerance technologies and experience.
\newblock In {\em International Symposium on Fault Tolerant Computing
  (FTCS-23)}, pages 2--9, 1993.

\bibitem{HuangK96}
Yennun Huang and Chandra M.~R. Kintala.
\newblock Software fault tolerance in the application layer.
\newblock In {\em Software Fault Tolerance}, pages 231--248. John Wiley \& Sons
  Ltd., Chichester, 1996.

\bibitem{HuangKKF95}
Yennun Huang, Chandra M.~R. Kintala, Nick Kolettis, and N.~Dudley Fulton.
\newblock Software rejuvenation: analysis, module and applications.
\newblock In {\em International Symposium on Fault Tolerant Computing
  (FTCS-25)}, pages 381--390, 1995.

\bibitem{IwamotoHMI02}
Chuzo Iwamoto, Tomonobu Hatsuyama, Kenichi Morita, and Katsunobu Imai.
\newblock Constructible functions in cellular automata and their applications
  to hierarchy results.
\newblock {\em Theor. Comput. Sci.}, 270(1--2):797--809, 2002.

\bibitem{JancarMPV95}
Petr Jancar, Frantisek Mr{\'a}z, Martin Pl{\'a}tek, and J{\"o}rg Vogel.
\newblock Restarting automata.
\newblock In {\em Fundamentals of Computation Theory (FCT 1995)}, volume 965 of
  {\em Lecture Notes in Comput. Sci.}, pages 283--292. Springer-Verlag, 1995.

\bibitem{Kappes99}
Martin Kappes.
\newblock Descriptional complexity of deterministic finite automata with
  multiple initial states.
\newblock {\em J. Autom. Lang. Comb.}, 5(3):269--278, 2000.

\bibitem{KappesK04}
Martin Kappes and Chandra M.~R. Kintala.
\newblock Tradeoffs between reliability and conciseness of deterministic finite
  automata.
\newblock {\em J. Autom. Lang. Comb.}, 9(2--3):281--292, 2004.

\bibitem{KappesKK00}
Martin Kappes, Chandra M.~R. Kintala, and Reinhard Klemm.
\newblock Formal limits on determining reliabilities of component-based
  software systems.
\newblock In {\em International Symposium on Software Reliability Engineering
  (ISSRE 2000)}, pages 356--364. IEEE Computer Society, 2000.

\bibitem{KappesN05}
Martin Kappes and Frank Nie{\ss}ner.
\newblock Succinct representations of languages by {DFA} with different levels
  of reliability.
\newblock {\em Theor. Comput. Sci.}, 330(2):299--310, 2005.

\bibitem{Kintala78}
Chandra M.~R. Kintala.
\newblock Refining nondeterminism in context-free languages.
\newblock {\em Math. Systems Theory}, 12(1):1--8, 1978/79.

\bibitem{Kintala83}
Chandra M.~R. Kintala.
\newblock Attributed grammars for query language translations.
\newblock In {\em Symposium on Principles of Database Systems}, pages 137--148.
  ACM, New York, 1983.

\bibitem{Kintala09}
Chandra M.~R. Kintala.
\newblock Software rejuvenation in embedded systems.
\newblock {\em J. Autom. Lang. Comb.}, 14(1):63--73, 2009.

\bibitem{KintalaF77}
Chandra M.~R. Kintala and Patrick~C. Fischer.
\newblock Computations with a restricted number of nondeterministic steps.
\newblock In {\em {S}ymposium on {T}heory of {C}omputing (STOC 1977)}, pages
  178--185. ACM, New York, 1977.

\bibitem{KintalaF80}
Chandra M.~R. Kintala and Patrick~C. Fischer.
\newblock Refining nondeterminism in relativized polynomial-time bounded
  computations.
\newblock {\em SIAM J. Comput.}, 9(1):46--53, 1980.

\bibitem{KintalaPW93}
Chandra M.~R. Kintala, Kong-Yee Pun, and Detlef Wotschke.
\newblock Concise representations of regular languages by degree and
  probabilistic finite automata.
\newblock {\em Math. Systems Theory}, 26(4):379--395, 1993.

\bibitem{KintalaW80}
Chandra M.~R. Kintala and Detlef Wotschke.
\newblock Amounts of nondeterminism in finite automata.
\newblock {\em Acta Inform.}, 13(2):199--204, 1980.

\bibitem{KintalaW86}
Chandra M.~R. Kintala and Detlef Wotschke.
\newblock Concurrent conciseness of degree, probabilistic, nondeterministic and
  deterministic finite automata.
\newblock In {\em Symposium on Theoretical Aspects of Computer Science ({STACS}
  1986)}, volume 210 of {\em Lecture Notes in Comput. Sci.}, pages 291--305.
  Springer-Verlag, 1986.

\bibitem{KleinK02}
Andreas Klein and Martin Kutrib.
\newblock Deterministic {T}uring machines in the range between real-time and
  linear-time.
\newblock {\em Theor. Comput. Sci.}, 289(1):253--275, 2002.

\bibitem{Knuth68}
Donald~E. Knuth.
\newblock Semantics of context-free languages.
\newblock {\em Math. Systems Theory}, 2(2):127--145, 1968.

\bibitem{Kutrib02}
Martin Kutrib.
\newblock Refining nondeterminism below linear time.
\newblock {\em J. Autom. Lang. Comb.}, 7(4):533--547, 2002.

\bibitem{Kutrib03}
Martin Kutrib.
\newblock Dimension- and time-hierarchies for small time bounds.
\newblock In {\em Fundamentals of Computation Theory ({FCT} 2003)}, volume 2751
  of {\em Lecture Notes in Comput. Sci.}, pages 321--332. Springer-Verlag,
  2003.

\bibitem{KutribM07a}
Martin Kutrib and Andreas Malcher.
\newblock Context-dependent nondeterminism for pushdown automata.
\newblock {\em Theoret. Comput. Sci.}, 376(1--2):101--111, 2007.

\bibitem{KutribM07}
Martin Kutrib and Andreas Malcher.
\newblock Finite turns and the regular closure of linear context-free
  languages.
\newblock {\em Discrete Appl. Math.}, 155(16):2152--2164, 2007.

\bibitem{KutribM09}
Martin Kutrib and Andreas Malcher.
\newblock Cellular automata with sparse communication.
\newblock In {\em Conference on Implementation and Application of Automata
  (CIAA 2009)}, volume 5642 of {\em Lecture Notes in Comput. Sci.}, pages
  34--43. Springer-Verlag, 2009.

\bibitem{KutribM09a}
Martin Kutrib and Andreas Malcher.
\newblock On one-way one-bit ${O}(1)$-message cellular automata.
\newblock {\em Electr. Notes Theor. Comput. Sci.}, 252:77--91, 2009.

\bibitem{KutribMW09}
Martin Kutrib, Andreas Malcher, and Larissa Werlein.
\newblock Regulated nondeterminism in pushdown automata.
\newblock {\em Theor. Comput. Sci.}, 410(37):3447--3460, 2009.

\bibitem{Leung98}
Hing Leung.
\newblock On finite automata with limited nondeterminism.
\newblock {\em Acta Inform.}, 35(7):595--624, 1998.

\bibitem{Leung98-2}
Hing Leung.
\newblock Separating exponentially ambiguous finite automata from polynomially
  ambiguous finite automata.
\newblock {\em SIAM J. Comput.}, 27(4):1073--1082, 1998.

\bibitem{Leung05}
Hing Leung.
\newblock Descriptional complexity of {NFA} of different ambiguity.
\newblock {\em Int. J. Found. Comput. Sci.}, 16(5):975--984, 2005.

\bibitem{Leung06}
Hing Leung.
\newblock Structurally unambiguous finite automata.
\newblock In {\em Conference on Implementation and Application of Automata
  (CIAA 2006)}, volume 4094 of {\em Lecture Notes in Comput. Sci.}, pages
  198--207. Springer-Verlag, 2006.

\bibitem{LeungW00}
Hing Leung and Detlef Wotschke.
\newblock On the size of parsers and {${\rm LR}(k)$}-grammars.
\newblock {\em Theoret. Comput. Sci.}, 242(1--2):59--69, 2000.

\bibitem{LiangCHKLTW04}
Deron Liang, P.~Emerald Chung, Yennun Huang, Chandra M.~R. Kintala, Woei-Jyh
  Lee, Timothy~K. Tsai, and Chung-Yih Wang.
\newblock {NT}-{SwiFT}: Software implemented fault tolerance on {W}indows {NT}.
\newblock {\em Journal of Systems and Software}, 71(1--2):127--141, 2004.

\bibitem{Malcher05}
Andreas Malcher.
\newblock On two-way communication in cellular automata with a fixed number of
  cells.
\newblock {\em Theoret. Comput. Sci.}, 330(2):325--338, 2005.

\bibitem{Malcher07}
Andreas Malcher.
\newblock On recursive and non-recursive trade-offs between finite-turn
  pushdown automata.
\newblock {\em J. Autom. Lang. Comb.}, 12(1--2):265--277, 2007.

\bibitem{MalcherMP08}
Andreas Malcher, Carlo Mereghetti, and Beatrice Palano.
\newblock Sublinearly space bounded iterative arrays.
\newblock In {\em Automata and Formal Languages (AFL 2008)}, pages 292--301.
  Hungarian Academy of Sciences, Budapest, Hungary, 2008.

\bibitem{MalcherP07}
Andreas Malcher and Giovanni Pighizzini.
\newblock Descriptional complexity of bounded context-free languages.
\newblock In {\em Developments in Language Theory (DLT 2007)}, volume 4588 of
  {\em Lecture Notes in Comput. Sci.}, pages 312--323. Springer-Verlag, 2007.

\bibitem{Mandl73}
Robert Mandl.
\newblock Precise bounds associated with the subset construction on various
  classes of nondeterministic finite automata.
\newblock Princeton Conference on System Sciences, 1973.

\bibitem{Masopust09}
Tom{\'a}s Masopust.
\newblock Regulated nondeterminism in {PDA}s: The non-regular case.
\newblock In {\em Non-Classical Models of Automata and Applications ({NCMA}
  2009)}, volume 256 of {\em books@ocg.at}, pages 181--194. {\"O}sterreichische
  Computer Gesellschaft, Wien, 2009.

\bibitem{MereghettiP00}
Carlo Mereghetti and Giovanni Pighizzini.
\newblock Optimal simulations between unary automata.
\newblock {\em SIAM J. Comput.}, 30(6):1976--1992, 2000.

\bibitem{MeyerF71}
Albert~R. Meyer and Michael~J. Fischer.
\newblock Economy of description by automata, grammars, and formal systems.
\newblock In {\em IEEE Twelfth Annual Symposium on Switching and Automata
  Theory}, pages 188--191. IEEE, 1971.

\bibitem{MrazPO07}
Frantisek Mr{\'a}z, Martin Pl{\'a}tek, and Friedrich Otto.
\newblock A measure for the degree of nondeterminism of context-free languages.
\newblock In {\em Conference on Implementation and Application of Automata
  (CIAA 2007)}, volume 4783 of {\em Lecture Notes in Comput. Sci.}, pages
  192--202. Springer-Verlag, 2007.

\bibitem{Niessner04}
Frank Nie{\ss}ner.
\newblock B{\"u}chi automata and their degrees of nondeterminism and ambiguity.
\newblock {\em J. Autom. Lang. Comb.}, 9(2--3):347--363, 2004.

\bibitem{Otto03}
Friedrich Otto.
\newblock Restarting automata and their relations to the {C}homsky hierarchy.
\newblock In {\em Developments in Language Theory (DLT 2007)}, volume 2710 of
  {\em Lecture Notes in Comput. Sci.}, pages 55--74. Springer-Verlag, 2003.

\bibitem{Otto06}
Friedrich Otto.
\newblock Restarting automata.
\newblock In {\em Recent Advances in Formal Languages and Applications},
  volume~25 of {\em Studies in Computational Intelligence}, pages 269--303.
  Springer-Verlag, 2006.

\bibitem{RavikumarI89}
Bala Ravikumar and Oscar~H. Ibarra.
\newblock Relating the type of ambiguity of finite automata to the succinctness
  of their representation.
\newblock {\em SIAM J. Comput.}, 18(6):1263--1282, 1989.

\bibitem{SalomaaWY94}
Kai Salomaa, Derick Wood, and Sheng Yu.
\newblock Pumping and pushdown machines.
\newblock {\em RAIRO Inform. Th\'eor. Appl.}, 28(3--4):221--232, 1994.

\bibitem{SalomaaY93}
Kai Salomaa and Sheng Yu.
\newblock Limited nondeterminism for pushdown automata.
\newblock {\em Bulletin of the EATCS}, 50:186--193, 1993.

\bibitem{SalomaaY94}
Kai Salomaa and Sheng Yu.
\newblock Measures of nondeterminism for pushdown automata.
\newblock {\em J. Comput. System Sci.}, 49(2):362--374, 1994.

\bibitem{SalomaaY97}
Kai Salomaa and Sheng Yu.
\newblock N{F}{A} to {D}{F}{A} transformation for finite languages over
  arbitrary alphabets.
\newblock {\em J. Autom. Lang. Comb.}, 2(3):177--186, 1997.

\bibitem{SchmidtS77}
Erik~M. Schmidt and Thomas~G. Szymanski.
\newblock Succinctness of descriptions of unambiguous context-free languages.
\newblock {\em SIAM J. Comput.}, 6(3):547--553, 1977.

\bibitem{StearnsH85}
Richard~E. Stearns and Harry~B. Hunt, III.
\newblock On the equivalence and containment problems for unambiguous regular
  expressions, regular grammars and finite automata.
\newblock {\em SIAM J. Comput.}, 14(3):598--611, 1985.

\bibitem{Sunckel98}
Bettina Sunckel.
\newblock {E}ine {H}ierarchie und ein {A}utomatenmodell f{\"u}r attributierte
  {G}rammatiken ohne geerbte {A}ttribute.
\newblock Master's thesis, J.W. Goethe-Universit{\"a}t Frankfurt, 1998.

\bibitem{Valiant76}
Leslie~G. Valiant.
\newblock A note on the succinctness of descriptions of deterministic
  languages.
\newblock {\em Information and Control}, 32(2):139--145, 1976.

\bibitem{VelosoG79}
Paulo A.~S. Veloso and Arthur Gill.
\newblock Some remarks on multiple-entry finite automata.
\newblock {\em J. Comput. System Sci.}, 18(3):304--306, 1979.

\bibitem{VermeirS81}
Dirk Vermeir and Walter~J. Savitch.
\newblock On the amount of nondeterminism in pushdown automata.
\newblock {\em Fund. Inform.}, 4(2):401--418, 1981.

\bibitem{WangHVCK95}
Yi-Min Wang, Yennun Huang, Kiem-Phong Vo, Pi-Yu Chung, and Chandra M.~R.
  Kintala.
\newblock Checkpointing and its applications.
\newblock In {\em International Symposium on Fault Tolerant Computing
  (FTCS-25)}, pages 22--31, 1995.

\bibitem{Wich00}
Klaus Wich.
\newblock Sublinear ambiguity.
\newblock In {\em Mathematical Foundations of Computer Science ({MFCS} 2000)},
  volume 1893 of {\em Lecture Notes in Comput. Sci.}, pages 690--698.
  Springer-Verlag, 2000.

\bibitem{Wich01}
Klaus Wich.
\newblock Characterization of context-free languages with polynomially bounded
  ambiguity.
\newblock In {\em Mathematical Foundations of Computer Science ({MFCS} 2001)},
  volume 2136 of {\em Lecture Notes in Comput. Sci.}, pages 703--714.
  Springer-Verlag, 2001.

\bibitem{Wich05}
Klaus Wich.
\newblock Sublogarithmic ambiguity.
\newblock {\em Theor. Comput. Sci.}, 345(2--3):473--504, 2005.

\end{thebibliography}

\end{document}